# ANTI PLAGIARISM APPLICATION WITH ALGORITHM KARP-RABIN AT THESIS IN GUNADARMA UNIVERSITY

**A. Benny Mutiara** and **Sinta Agustina**
Graduate Program in Information System
Gunadarma University
Depok, Indonesia
Email : amutiara@staff.gunadarma.ac.id

## Abstracts

*Plagiarism that is plagiarizing or composition retrieval, opinion, etcetera from other people and makes it is likely composition and opinion him-self. Plagiarism can be considered to be crime because stealing others copyrights. Like action a student copying some part of writings without valid permission from the original writer. In education world, plagiarism perpetrator can get the devil to pay from school/university. Plagiarism perpetrator conceived of plagiator. This thing is possible unable to be paid attention by the side of campus because of limitation from some interconnected factors for example student amounts Gunadarma University reaching thousands and incommensurate to tester amounts or lecturer the side of campus in charge directs problem thesis. In this paper, an application have been developed in order to check and look for 5 type percentage similarity from a thesis with other one at certain part or chapters. Percentage got that is 0%, under 15%, between 15-50%, up to 50% and 100%. So it should be expected that the results could be used by thesis advisor and also thesis examiner from the Student at Gunadarma University.*

## I. INTRODUCTION

Along with growing it information technology and computerization that is increasingly fast, hence production process a writing masterpiece can be done easily and quickly. For example a student is being does final duty at institution of certain education. Research done must fulfill scientific procedure, like identification of problem, formulation of hypothesis, usage of research method, and the analysis framework. To do it all, required ability of adequate academic. But not all students has ability of academic enough to compile end duty, even blaze with research. They are this till now many finding difficulties and finally chooses short cut by doing plagiarism.

Intention of this writing is to develop application that should be applied to look for some equality gratuities from a part at a thesis with other one.

## II. BASIS THEORY

### 1. Plagiarism

Plagiarism is plagiarizing or composition retrieval, opinion, etcetera from other people and makes it is likely composition and opinion him-self [1].

Classified of plagiarism [2]:
- applies others article rawly, without giving sign to explain ( for example by using different quotation mark or paragraph block) that the text is taken precisely from other article
- takes others idea without giving annotation that is enough about the source

### 2. Copyrights

Copyrights (international symbol: ©) it is exclusive rights of Creator or Right Owner Creates to arrange usage result of infusion of idea or certain information. Basically, a copyright is "the right to copy a creation". A copyright earns also enables the right owner to limit illegal duplication to a creation. In general also, copyrights has a period of applying is certain ness limited.

### 3. PHP

PHP is language (scripting language) what designed peculiarly for usage at Website. PHP is tool for making of page Website which dynamic. Rich feature would make design of

Website and easier programming, PHP applied by 13 million domains (according to survey Netcraft at www.php.net/usage.php). PHP abbreviation for HyperText Preprocessor. In the early of it's the expansion by Rasmus Lefdorf, he calls it as tools Personal Home Page.

### 4. My SQL

MySQL is one of server database that is enough to be recognized the existing. MySQL output T.c.X.Data ConsultAB, a company IT Swedia, what offers various excellences compared to other server database [3]:
1. Can handle millions of user during is at the same
2. Can support more than 50.000.000 records
3. Hardly quickly in executing comand

### 5. XML

XML is a technology cross platform, and is tool to do information transmission. XML is not program, or book. XML is a technology, a standard with various certain orders. In understanding that is simple, a document XML only ordinary texts file comprising various tag defined itself by XML documents maker. As according to the name, eXtensible Markup Language, a document XML is a document with markup, same as the things of with HTML. But, XML is not design to replace HTML. XML is more designed description of data to and focused itself at the data.

### 6. String Matching Algorithm

Matching of String consisted of stages steps to find one string or more to all case at a string ( in general is called as pattern string) in text. All matching algorithm of string will yield all pattern string found on text

**Karp-Rabin Algorithm**
At repair of method brute force can be classified to follow the sequence comparison of pattern character and character text for every attempt. At the comparison process there are four categories [4]:
1. From right to left
2. From left to right-
3. In specific order
4. In any order

Based on four above categories, algorithm Karp-Rabin included into category from left to right. Algorithm Karp-Rabin applies function of hash providing simple method to avoid time complexity $O(m^2)$. Than checked position every pattern which there is in text, would more efficiently if done only at pattern wanted. Equality checking between two words applies function of hash.

Function of hash must have propertys as follows [4]:
1. Ability of efficient computing
2. High discrimination to string
3. Function of hash ( y [ j+1 .. j+m ] ) must easy to be computing from
   - hash ( y[j .. j+m-1] )
   - hash ( y[j+m] )

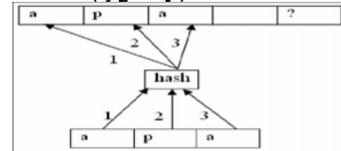

**Fig.1.** Illustration of Karp-Rabin Algorithm [4]

Algorithm Karp-Rabin has marking as follows:
- Applies function of hash
- Phase preproses in time complexity O(m) and constant place..
- Seeking phase in time complexity O(mn)
- O(n+m) estimates active time

Function of hash also applied default value index to or key and applied then each time data relating to value or key is taken. In a seeking, hence first time name would be hashing with function of the same hash when save the data (index) causing yields a value which will be compared to at data is index with the value. Hence in general seeking with 10 possibilities ( digit 0-9) would be quicker compared to based on 26 possibilities ( character a-z)

**Declaration hash function of this apllication**

*Name file :add_file_ to_sql.php*

$xml.="\t<sentence sentence_id=\"$j\">";

**$hash=my_hash**(strtolower(remove_un($line)));

$text= htmlspecialchars($line,ENT_QUOTES);

//add_line_sql(**$hash**,$id_doc['id'],$i,$j);

mysql_query("INSERT INTO text

(id,hash,id_doc,id_para,id_phrase)

VALUES('','".**$hash**."','".$id_doc['id']."','".$i."','".$j."')") or

die(mysql_error());

There is first declaration where data at every file will be encryption into form of this number and hash value will be kept in SQL

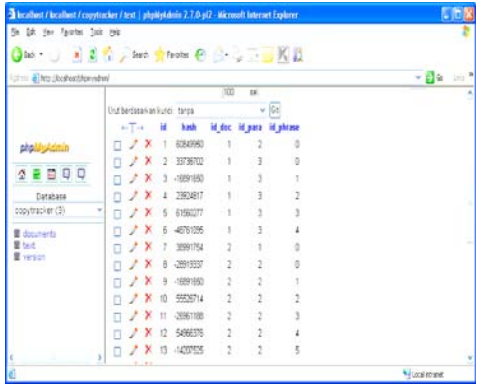
**Fig.2.** Internal Storage SQL

Then result of its, will be sent into form of xml. As a sentence_id.

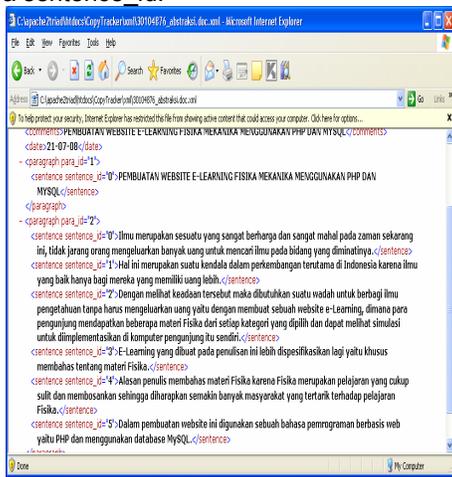
**Fig.3.** Declaration at form of xml

### III. RESEARCH METHODOLOGIES

1. **Field Study and literature**
   Field study is done by the way of retrieval of data directly at Univeristy's library.

2. **Scheme of navigation structure**
   Navigation structure depicts visually stepped or road (street sequence an application.

3. **System scheme**
   Scheme of system at the application of this applied by using Unified Modeling Language (UML) applied to design model a system. Scheme of system model at the application of this depicted through Use Case Diagram and Activity Diagram.

4. **Scheme of design application**
   Scheme of interface from application aim to design a page beforehand then its implementation to coding and give a picture or story board about application is being designed.

5. **Making the code**
   At this process writer looks for some coding or syntax related to source of internet media where in it have been there is link with the data base.

6. **Testing**
   Testing process is done by after upload process at files indication.

### IV. ANALYSIS AND DISCUSSION

**1 Starts application**

At this step, writer will test anti plagiatism application built applies PHP. Before all writer will give initial picture of usage of application before user does documents comparison process. Process first time done to implement application that is :
- Install apache2triad514.exe
- Puts down one folders so called CopyTracker in htdocs like picture:

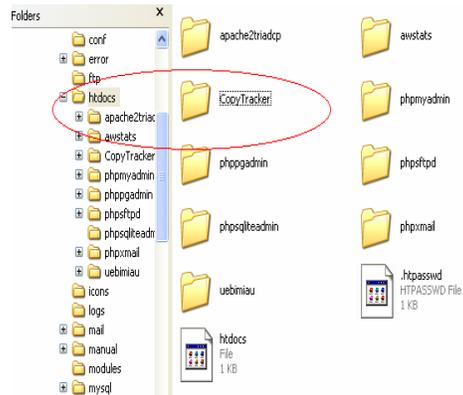
**Fig.4.** Folder CopyTracker in htdocs.

- Then user types
  http://localhost/CopyTracker/ at browser then will seen like at picture following :

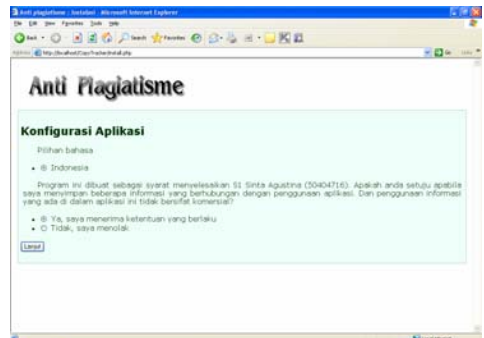
**Fig.5.** Installation process of Application

Then application in default will open Home page like at picture following :

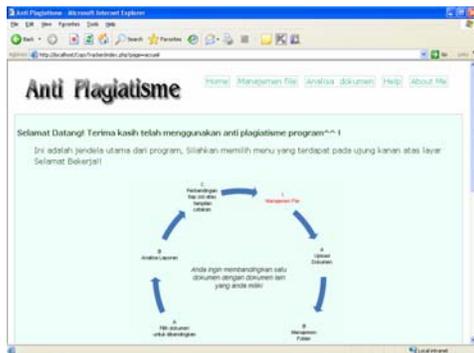

**Fig.6.** Home Display after installation process

**2. Upload and chooses file which will be compared**

After user clicks management of file hence user can add documents into documents data base. With clicking sign 'browse' to choose file which wish to be compared

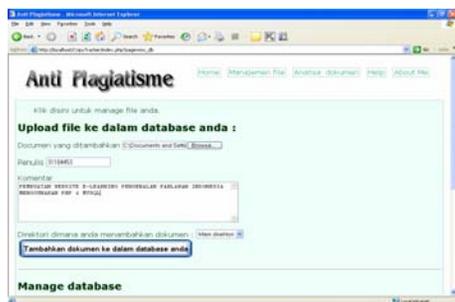

**Fig.7.** Example of file upload process into data base

If upload process have succeeded carefully hence automatically program will inform the program has successfully is upload.

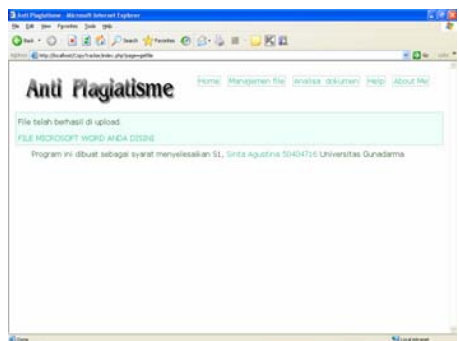

**Fig.8.** File uploads successfully

**3. Application testing**

***a. Examination of file is indication 0%***

Examination to file indication by 0 % is examination where two files compared to really differing in in contents and formation of sentence as a whole.
Example of at 30104599-abstraksidoc with 50404783-abstraksidoc.

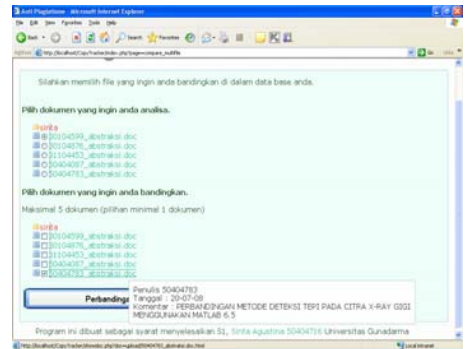

**Fig. 9.** File Comparison

Result is following :

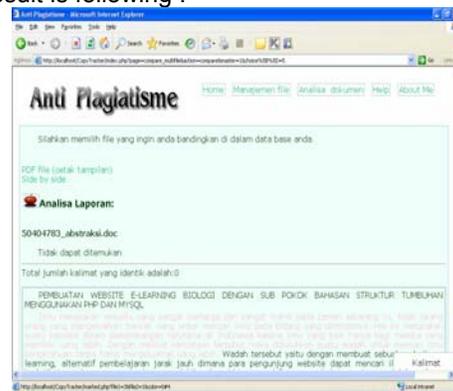

**Fig.10.** Files is indication 0% success detected

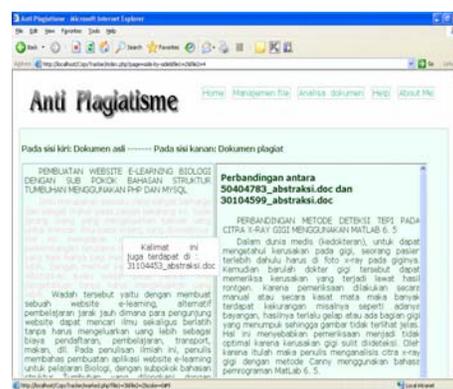

**Fig.11.** Display by test result indication 0 %

***b. Examination of file is indication under 15%***

Examination to file indication less than 15 % is examination where two files compared to has equality which can be told a few. Example of at

files 30104876-abstraksi.doc and 31104453-abstraksi.doc.

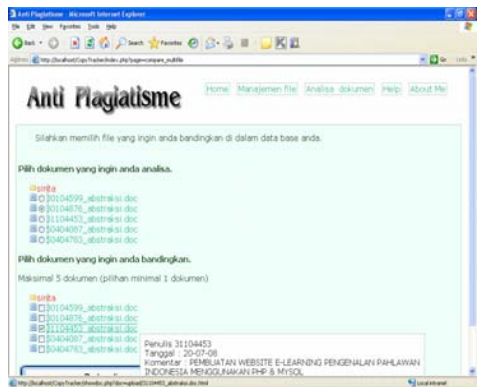

**Fig.12.** File comparison

Result is following :

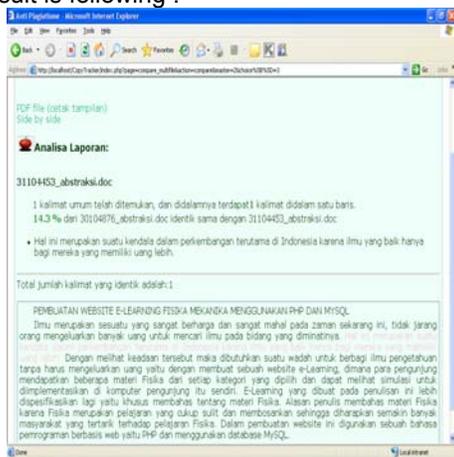

**Fig. 13.** Files are indication under 15% success detected

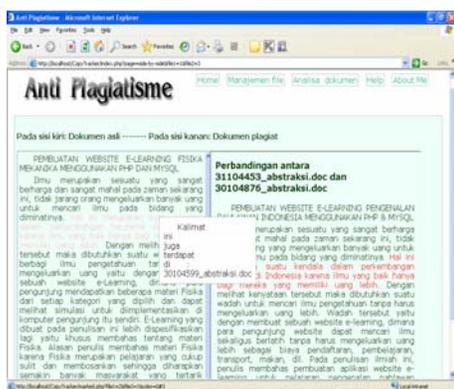

**Fig.14.** Display by test indication under 15 %

Some the same sentences visible at report analysis under this and word having equality will be boldface. From application analysis, at both this files there is equality at 1 of identical sentence from a paragraph containing 7 sentence in file 31104453-abstraksi.doc is coming from 30104876-abstraksi.doc Causing is obtained calculation 14,3 % identical sentence. Gotten from 1 : 7 x 100% = 14,285% and rounded up to to become 14,3%. Besides at file 30104876-abstraksi.doc which is original file also there is a sentence indication is same at file 30104599-abstraksi.doc because this file also there is by one the same data bases.

### c. Examination of file is indication between 15-50%

Examination to file indication between 15-50 % is examination where two files compared to has equality which can be told stays at midpoint an action of plagiarism. Example of at file 30104599-abstrakasi.doc and 31104453-abstraksi.doc

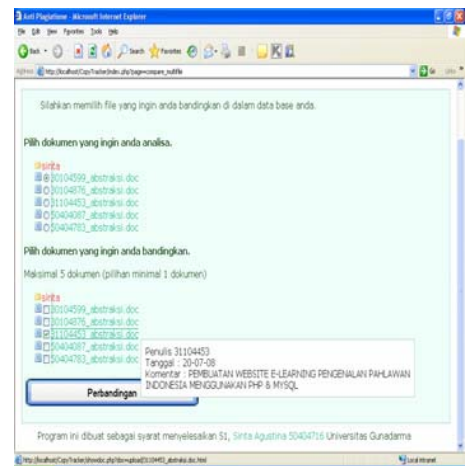

**Fig.15.** File comparison

Result is following:

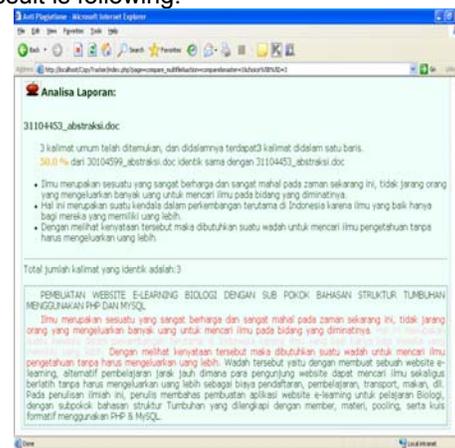

**Fig.16.** Files is indication between 15-50 % success detected

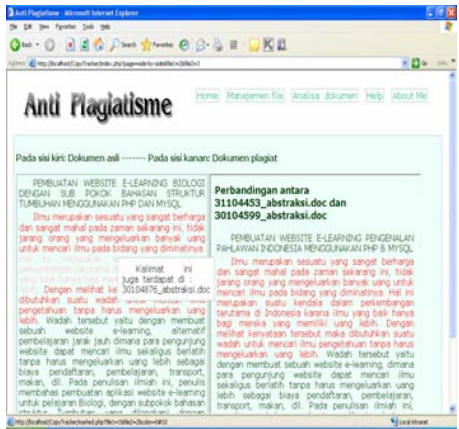

**Fig.17.** Displays by test result indication between 15-50 %

Some the same sentences visible at report analysis under this and word having equality will be boldface. From application analysis, at both this files there is equality at 3 of identical sentence from a paragraph in file file 31104453-abstraksi.doc. is coming from Full scale 30104599-abstraksi.doc of sentence found on one paragraphs is 6. So is obtained calculation counted 50% identical sentence. Gotten from 3 : 6 x 100% = 50% Besides at file 30104599-abstraksi.doc which is original file also there is a sentence indication is same at file 30104876-abstraksi.doc because this file also there is by one the same data bases and given to ruddles young. While at file 31104453-abstraksi.doc sentence indication by plagiarism is given to ruddles.

***d. Examination to file indication up to 50%***
Examination to file indication up to 50% is examination where two files compared to has equality which can be told stays at upper point and comes near action of plagiatism. Example of at file 50404783-abstraksi.doc and 50404087-abstraksi.doc

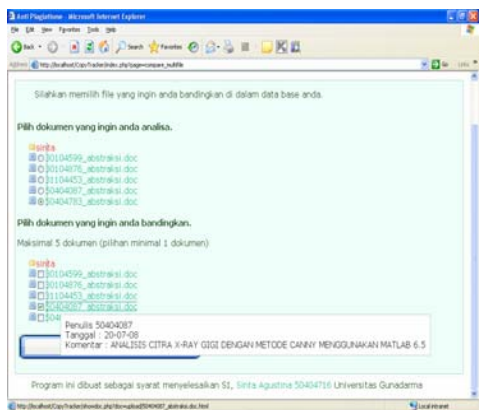

**Fig.18.** File comparison

Result is following:

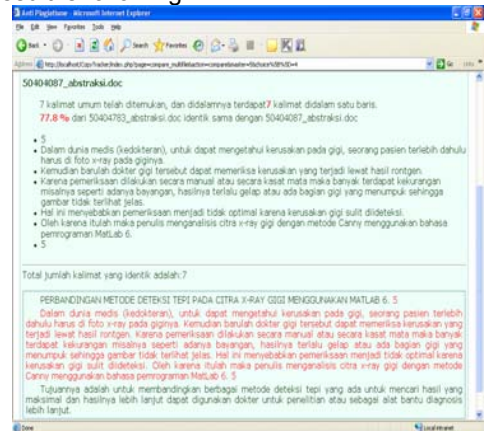

**Fig. 19.** Files is indication up to 50% success detected

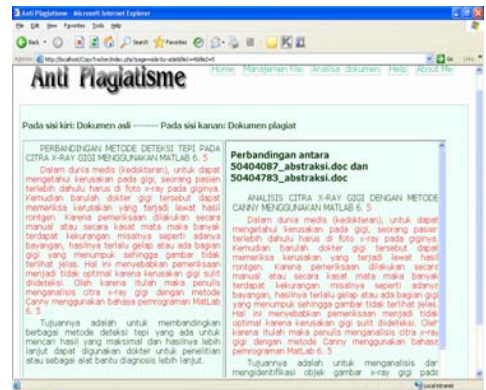

**Fig. 20.** Displays by test result indication up to 50 %

Some the same sentences visible at report analysis under this and word having equality will be boldface. From application analysis, at both this files there is equality at 7 of identical sentence from file 50404087-abstraksidoc. is coming from Full scale 50404783-abstraksidoc of sentence found on one paragraphs is 9. So is obtained calculation 77,8 % identical sentence. Where gotten from 7 : 9 x 100% = 77,77778 rounded up to to become 77,8 %

***e. Examination to file indication 100%***

Examination to file indication between 100% is examination where two files compared to has equality at its contents as a whole from beginning of finite of akhirContoh applied at this examination is 50404783-abstraksi.doc with 50404783-abstraksi.doc Logically has can be told that both this files is same in consisting in writing format in it.

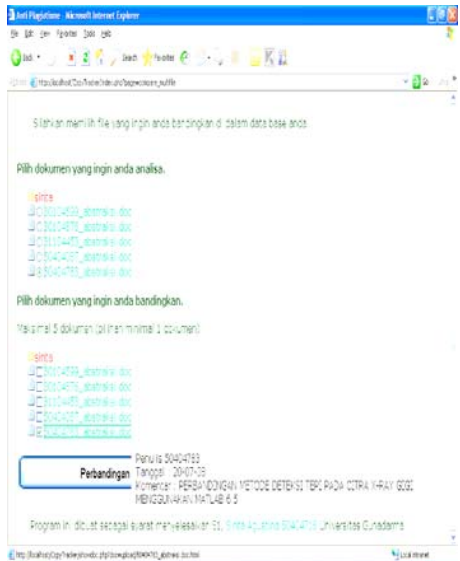

**Fig.21.** File comparison

Result is following :

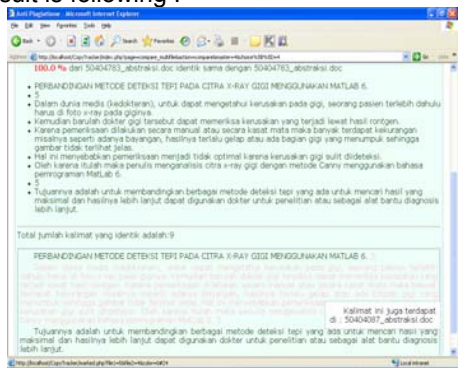

**Fig. 22**. File is indication 100 % success detected

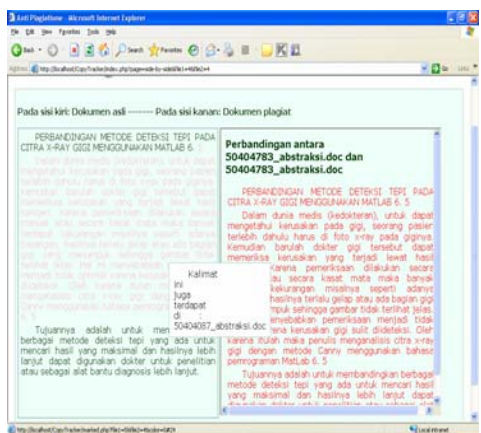

**Fig. 22.** Displays by test result up to 50 %

The same sentence visible at report analysis under this and word having equality will be boldface. From application analysis, at both this files there is equality at 9 of identical sentence from a paragraph in file 50404783-abstraksi.doc is coming from Full scale 50404783-abstraksi.doc of sentence found on one paragraphs is 9. So is obtained calculation 100 % identical sentence. This thing is because compared to be two the same file. But at this file also is indication equal to file 50404087-abstraksi.doc because residing in at the same data base so that sentence is indication color to pink.

## V. CONCLUSION

This anti plagiarism application is expected to be applicable at the education world. This application is useful if a lecturer wish to verify duty of the students and also for area of larger ones like at processing thesis. The aim of this application is to look for percentage of similarity two the files. From percentage results it could be justified at any boundary someone could be told as a plagiator.

[14.] anonim. http://sourceforge.net/